\documentstyle[12pt,epsf,axodraw,cite]{article}
\def\be{\begin{equation}}
\def\ee{\end{equation}}
\def\barr{\begin{array}}
\def\earr{\end{array}}
\def\dis{\displaystyle}

\def\gev{\; {\rm GeV} }
\def\bea{\begin{eqnarray}}
\def\eea{\end{eqnarray}}

\def\bold#1{\setbox0=\hbox{$#1$}
     \kern-.025em\copy0\kern-\wd0 
     \kern.05em\copy0\kern-\wd0 
     \kern-.025em\raise.0433em\box0 }

\def\etal{ {\em et al.}}

\begin{document}
\vspace*{-1in}
\renewcommand{\thefootnote}{\fnsymbol{footnote}}
\begin{flushright}
MRI-P-010302\\
\texttt{hep-ph/0102199} 
\end{flushright}
\vskip 5pt
\begin{center}
{\Large{\bf Muon anomalous magnetic moment confronts 
exotic fermions and gauge bosons}}
\vskip 25pt
{\sf Debajyoti Choudhury 
\footnote{E-mail address: debchou@mri.ernet.in}}, 
{\sf Biswarup Mukhopadhyaya 
\footnote{E-mail address: biswarup@mri.ernet.in}},   
and
{\sf Subhendu Rakshit 
\footnote{E-mail address: srakshit@mri.ernet.in}}  
\vskip 10pt 
{\em Harish-Chandra Research Institute, Chhatnag Road, Jhusi, Allahabad
211 019, India }\\

\vskip 20pt

{\bf Abstract}
\end{center}

\noindent
\begin{quotation}
  {\small We investigate the status of models containing exotic fermions or
    extra $Z$-like neutral gauge bosons in the light of the recent data on
    anomalous magnetic moment of muon. We find that it is possible to extract
    interesting bounds on the parameters characterizing such models. The 
    bounds are particularly strong if the new flavour-changing neutral 
    currents are axial vectorlike.
\vskip 10pt
\noindent   
} 
\end{quotation}

\vskip 20pt  

\setcounter{footnote}{0}
\renewcommand{\thefootnote}{\arabic{footnote}}
%
%
The recent measurement of the anomalous magnetic moment ($a_{\mu}$) of
the muon by the E821 experiment~\cite{BNL} at BNL may very well act as
an Occam's razor in constraining physics beyond the Standard Model
(SM). The data seem to indicate a $2.6\sigma$ deviation from
theoretical predictions based on the SM.  More precisely, the measured
value $a_\mu \equiv (g_\mu - 2)/2$ lies in the range
\be
a_\mu^{\rm exp} = (11\ 659\ 202\, \pm 14 \pm 6) \times 10^{-10}\nonumber
\ee
in units of the Bohr magneton ($e / 2 m_\mu$).  Accumulation of fresh data is
likely to reduce the errors even further. Comparing this with the SM
prediction~\cite{smvalue} of 
\be
a_\mu^{\rm SM} = (11\ 659\ 159.7 \pm 6.7) \times 10^{-10},\nonumber
\ee
one has a rather tantalizing allowance for contributions from a number
of non-standard options~\cite{mar,other,otherold,review}. At the same
time, it is possible to constrain new theories with potential extra
contributions to $(g_\mu-2)$. This is especially true, as is evident
from equations (1) and (2), for scenarios where the additional
contributions to the magnetic moment can be negative. One such set of
possibilities consists of models with exotic fermions and/or extra
neutral gauge bosons~\cite{Morris}. In this note we shall examine
constraints on parameters of these models, such as the couplings and
the masses of these extra particles, that can be imposed from the new
measurement of the muon anomalous magnetic moment.

\section{Exotic Leptons}
\begin{figure}[h]
\begin{center}
\begin{picture}(300,108)(0,-8)
\SetWidth{1.2}
\ArrowLine(0,100)(30,100)
\Text(5,107)[rb]{$\footnotesize{\mu}$}
\ArrowLine(100,100)(130,100)
\Photon(65,65)(65,18){2}{4}
\Text(70,40)[lb]{$\footnotesize{\gamma}$}
\Text(125,107)[lb]{$\footnotesize{\mu}$}
\Photon(100,100)(30,100){2}{4}
\Text(65,107)[cb]{$\footnotesize{Z}$}
\ArrowArc(65,100)(35,180,270)
\ArrowArc(65,100)(35,270,0)
\Text(36,70)[ct]{$\footnotesize{F}$}
\Text(93,70)[ct]{$\footnotesize{F}$}
\Text(65,12)[ct]{$\footnotesize{(a)}$}

\ArrowLine(180,100)(210,100)
\Text(185,107)[rb]{$\footnotesize{\mu}$}
\Photon(280,100)(210,100){2}{4}
\ArrowLine(280,100)(310,100)
\Photon(245,65)(245,18){2}{4}
\Text(250,40)[lb]{$\footnotesize{\gamma}$}
\Text(305,107)[lb]{$\footnotesize{\mu}$}
\Text(245,107)[cb]{$\footnotesize{Z'}$}
\ArrowArc(245,100)(35,180,270)
\ArrowArc(245,100)(35,270,0)
\Text(274,70)[ct]{$\footnotesize{\tau}$}
\Text(217,70)[ct]{$\footnotesize{\tau}$}
\Text(245,12)[ct]{$\footnotesize{(b)}$}
\end{picture} 
\end{center}
\vspace*{-6ex}
      \caption{\em Feynman diagrams contributing to $a_\mu$.
        Diagram {\em (a)} corresponds to theories with exotic 
        fermions while diagram {\em (b)} appears in theories with an 
        exotic $Z'$.}
        \label{fig1}
\end{figure}
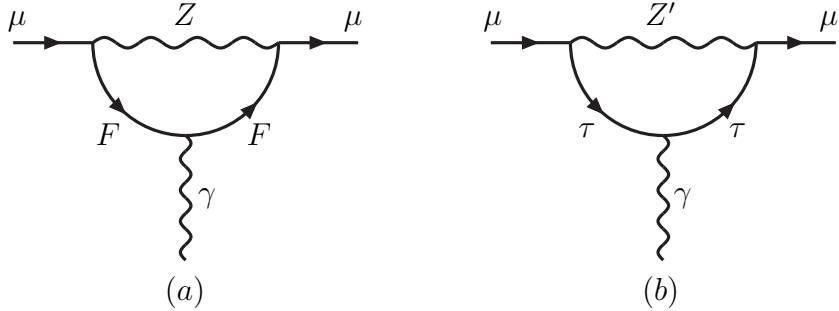

Let us first discuss models with exotic leptons. On demanding that
they have SM gauge interactions with the usual SM leptons, our choice
gets restricted to vectorlike doublets or singlets and mirror
fermions. For the last named ones, it may be noted that anomaly
cancellation requires the presence of entire mirror families
consisting of quarks and leptons. Constraints on ordinary--exotic
fermion mixing from electroweak observables have been already
discussed in the literature~\cite{Langacker}.  For example, LEP
results have imposed a model independent lower bound of $93.5\gev$ on
the mass of a charged exotic lepton~\cite{PDG}. A characteristic
feature of all these scenarios is the existence of flavour-changing
couplings of the $Z$-boson with fermions either in the left or in the
right-handed sector (or both).  These exotic fermions contribute to
muon anomalous magnetic moment via one loop diagrams shown in
Fig.~\ref{fig1}(a) where we generically represent the exotic lepton by
$F$.

Parametrizing the flavour changing neutral current vertex by
\be \Gamma^\mu_{\rm FCNC} = \frac{g}{2\cos\theta_W}\;
        \gamma^\mu(a_L P_L + a_R P_R),
\label{eq1} 
\ee 
it is a straightforward exercise to calculate the additional
contribution to $a_\mu$.
Under the simplifying assumption that only one species of such exotic
fermion ($F$) exists, we have~\cite{leveille}
\be
\barr{rcl}
a_{\mu} &=& \dis 
\left(\frac{g}{2\pi \cos \theta_W}\right)^2 \;\frac{m_\mu}{M_Z}\; r
    \;\Bigg[ (a_R + a_L)^2 \left\{  \frac{m_\mu}{M_Z} \; f_1 (r) 
                              + \frac{m_F}{M_Z} \; f_2 (r) 
                       \right\}
        \\[2ex]
& & \dis \hspace*{10em}
        + (a_R - a_L)^2 \left\{  \frac{m_\mu}{M_Z} \; f_1 (r) 
                              - \frac{m_F}{M_Z} \; f_2 (r) 
                       \right\}
   \Bigg]
           \\[2ex]
f_1(x) & = & \dis
\left(\frac{5}{6}-\frac{5}{2}x + x^2
        +(x^3-3x^2+2x)\; \ln\frac{x-1}{x}\right) 
           \\[2ex]
&+ & \dis \frac{x - 1 }{2 x} \: 
      \left(\frac{5}{6}+\frac{3}{2}\;x+x^2 + (x^2+x^3)
  \; \ln\frac{x-1}{x}\right)
         \\[3ex]
f_2 (x) & = & \dis \left(2x-1+2(x^2-x) \ln\frac{x-1}{x}\right)
+ 
\frac{1 - x}{2 x} \: \left(\frac{1}{2}+x+x^2
  \;\ln\frac{x-1}{x}\right)
\earr
        \label{formulae}
\ee
with $r\equiv (1-m_F^2/M_Z^2)^{-1}$.  

It should be mentioned here that the presence of exotic fermions also
implies flavour-changing Yukawa couplings. Therefore, additional
contributions accrue from Higgs-mediated loops as well.  Numerically,
however, these are relatively less significant, particularly for large
Higgs masses.

\begin{figure}[htb]
\vspace*{-2.5cm}
\centerline{
\epsfxsize=8.0cm\epsfysize=10.0cm
\epsfbox{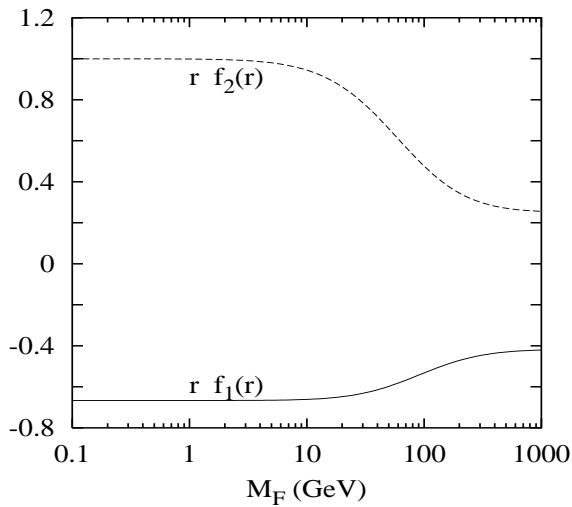} 
}
\caption{\em The functions $r f_{1,2}(r)$ that appear in 
        eqn.(\protect\ref{formulae}) for $r = (1 - m_F^2 / m_Z^2)^{-1}$.}
      \label{fig:the_fs}
\vspace*{-0.2cm}
\end{figure}

The functions $r f_{1,2}(r)$, exhibited in Fig.~\ref{fig:the_fs} are
seen to assume constant asymptotic values for both very small and very
large $r$.  Since these functions themselves are of the same order
(with $r f_2(r)$ actually being the larger of the two), clearly the
term in eqn.(\ref{formulae}) proportional to $m_F$ would tend to
dominate, particularly because we are considering $m_F$ on the order
of the electroweak scale or above. This ceases to be the case if
$a_L=0$ or $a_R=0$. ({\it i.e.} for pure left- or right-handed
couplings). In such cases, the chirality structure of the interaction
term itself rules out any non-zero contribution corresponding to a
mass insertion in the internal fermion line. Consequently, any
enhancement due to large $m_F$ is at best logarithmic and the ensuing
constraints from $a_\mu$ are not significant.

\begin{figure}[htb]
\vspace*{-2.0cm}
\hspace*{-0.5cm}
\centerline{
\epsfxsize=7.0cm\epsfysize=9.0cm
\epsfbox{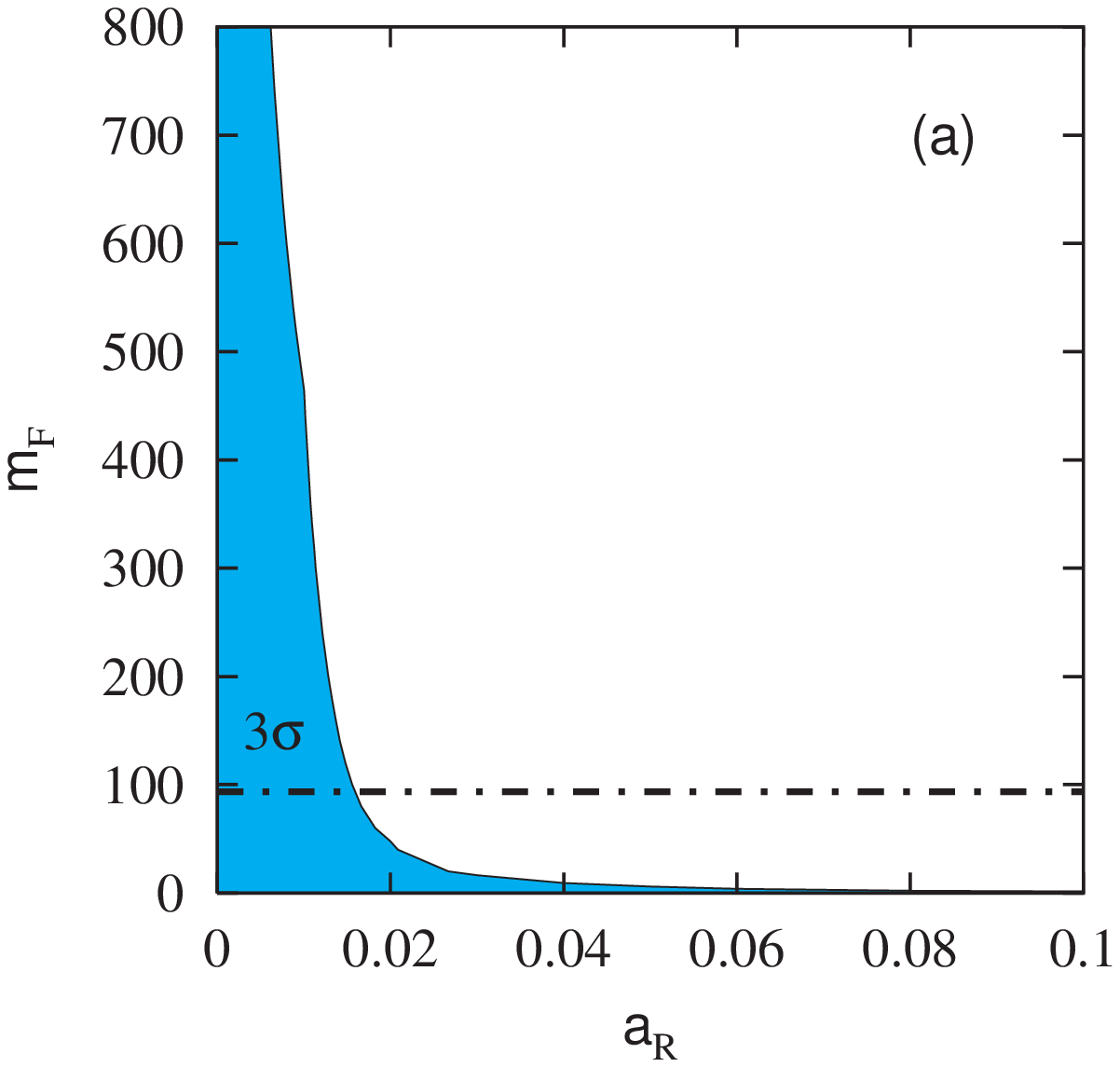} 
\vspace*{-0.0cm}
\hspace*{-0.cm}
\epsfxsize=7.0cm\epsfysize=9.0cm
\epsfbox{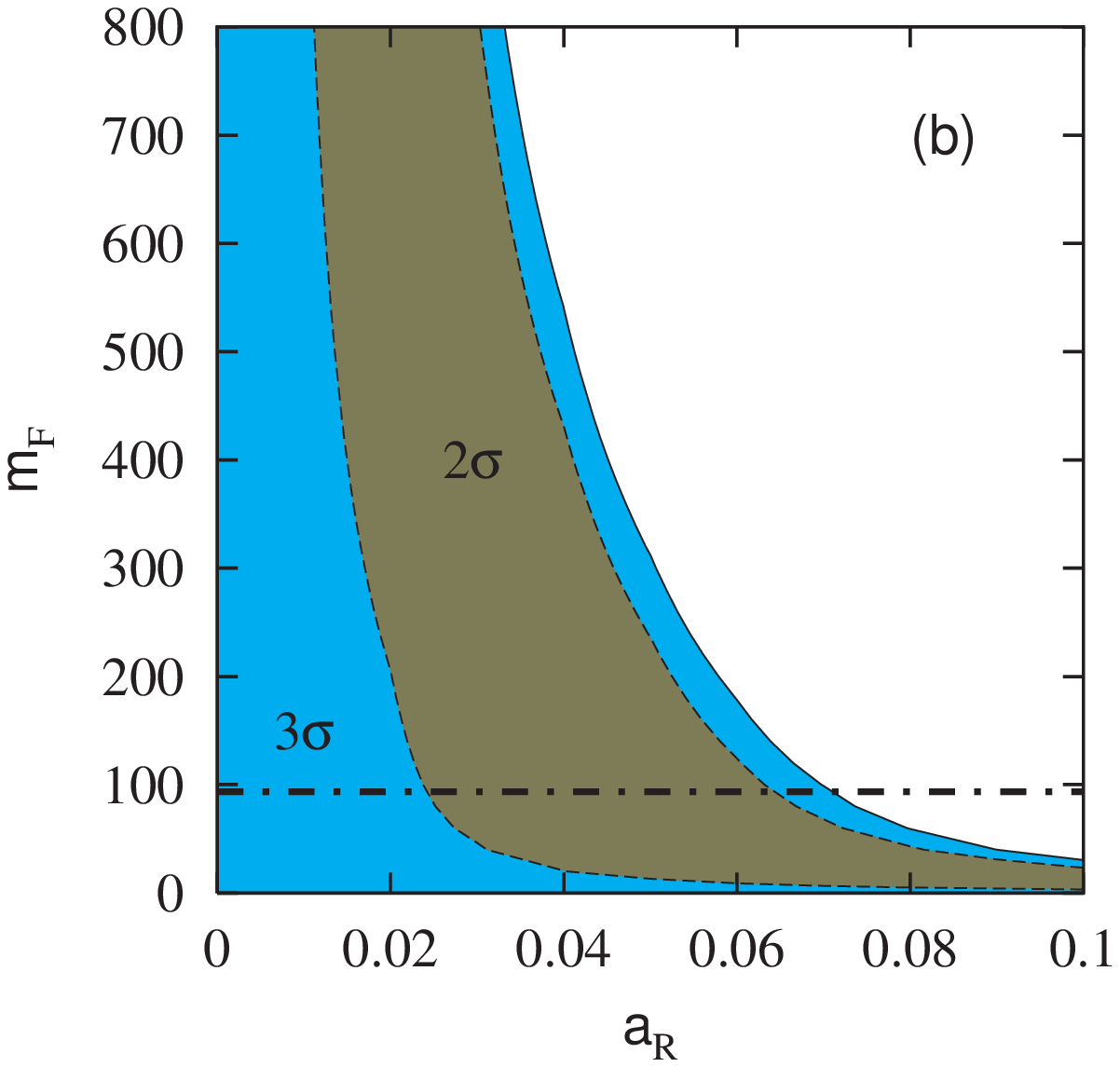}
\vspace*{-2.0cm}
}

\caption{\em The region of parameter space allowed by the $a_\mu$ data at
  different confidence levels for models containing an exotic lepton.
  The light-shaded region is allowed at the $3\sigma$ level, whereas
  the darker one is allowed at $2\sigma$ level. 
The region below the dash-dotted line is ruled out by LEP data~\protect\cite{PDG}. 
Fig.~{\em (a)}
  corresponds to $a_L=-a_R$, and while Fig.~{\em (b)}, to $a_L=a_R$.
  Case~{\em (a)} is inconsistent with the data at the $2\sigma$
  level.} 
      \label{extraf}
\end{figure}

The situation is more complicated for the general case where both
left- and right-handed couplings are present. In such a case, one has
to constrain a 3-dimensional parameter space spanned by ($m_F, a_L,
a_R$). We prefer to illustrate the issues involved by considering the
two extremes, namely, $a_L = \pm a_R$. The results for a more general
ratio between the two couplings lie in between the two extremes and
can be deduced by considering an appropriate combination of the same.
For $a_L=-a_R$ (a purely axial coupling), the additional contribution
is negative. Since the data allows for only a very small range for
$\delta a_\mu \equiv a_{\mu}^{\rm exp} -a_{\mu}^{\rm SM}<0$, ({\it
  i.e.}  for values of the $a_{\mu}$ lower than the SM prediction),
this immediately translates to a significant constraint in the
$a_R$--$m_F$ plane (see Fig.~\ref{extraf}). On the other hand,
$a_L=a_R$ (vectorlike current) leads to a positive contribution to
$a_\mu$.  Since the available allowance here is much larger, the
resulting constraints are weaker.

For example, for $a_L=-a_R$ with an exotic fermion mass of $400\gev$,
the maximum value of $a_R$ turns out to be about $1.0\times10^{-2}$.
Reverting to Fig.~\ref{fig:the_fs}, note that the function $r f_2(r)$
assumes the constant value of $1/4$ for large $m_F$. In the expression
for $a_\mu^{\rm exotic}$, this asymptotic value is multiplied though
by $m_F / m_Z$. Hence, the exotic fermion contribution to $a_\mu$
grows with larger $m_F$.  This is but a reflection of the
non-decoupling character of the exotic sector~\cite{appel} so long as
we do not correlate the exotic fermion mass and its mixing angle with
the ordinary fermion(s) but rather treat the two as independent
phenomenological parameters.  The tightening of the bound on $a_R$
with increasing $m_F$ is amply demonstrated by Fig.~\ref{extraf}.

If we examine specific models~\cite{E6} like those based on the $E_6$
gauge group, some additional features may reveal themselves. The
fundamental (27-dimensional) representation of $E_6$ contains 16, 10
and 1-plets of $SO(10)$. The 10-plet of $SO(10)$ contains a vectorlike
lepton isodoublet $(N_E, E)$ and the 1-plet is an isosinglet neutrino
$N$. In addition, the 16-plet contains a right-handed neutrino
($\nu_R$) which is singlet under the entire electroweak gauge group.
In this case there is an extra contribution coming from another
diagram where the photon couples to two W bosons and the $N_E$ sits
inside the loop. The corresponding contribution has a sign opposite to
that of the case we considered earlier and can be sizable. As a result
the total additional contribution to $a_\mu$ may turn out to be
positive, especially if we take the $W\overline{N}_E\mu$ coupling to
be the same as $Z\overline{F}\mu$ coupling.  However, this is not the
case in general, since $\nu_\mu$ could mix with all three of $\nu_E$, $N$ 
and $\nu_R$.  This extra mixing in the neutrino sector prevents an
exact or near-exact cancellation between the charged current loop and
the corresponding $Z$-mediated loop. Thus, if one has non-decoupling
contributions from both $Z$ and $W$ loops, the constraint from
$(g_\mu-2)$ has to be on a parameter space of larger dimensionality.
Any projection on the two dimensional subspace spanned by $(a_R,
m_\mu)$ would be weaker than those in Fig.~\ref{extraf}.

\section{Extended gauge sector}
Another set of potentially interesting contributions come from diagrams with
an extra $Z$-type boson ($Z'$). The breaking of $E_6$ results in more than one
additional $U(1)$ symmetries, at least one of which may survive down to the
TeV scale in many scenarios, leading to an extra neutral gauge boson of
phenomenological significance.  The ensuing contributions to $(g_\mu-2)$ will
involve a graph similar to the one shown in the previous figure, but with the
muon replacing $F$, and $Z'$ replacing $Z$. As has already been
shown~\cite{mar}, the smallness of the muon mass in the loop makes the
contribution rather small, and one hardly obtains a bound of any significance.

If however, one considers a model where the $Z'$ has {\it
  flavour-changing couplings} to the known charged leptons, then the
data on $a_\mu$ can subject it to more stringent limits. The
constraints on the mass of such a boson is model-dependent~\cite{PDG}.
For example if one extends the SM gauge group by an additional $U(1)$,
with respect to which the $\mu$ and the $\tau$ have different charges, 
then the corresponding $Z'$ has an unsuppressed gauge coupling with a
$\bar\mu \tau$ current~\cite{gautam}.  In such cases, one has
contributions to the muon anomalous magnetic moment via one-loop
diagram in Fig.~\ref{fig1}(b). Once again eqn.(3) gives the generic
structure of the flavour changing neutral current vertex.

\begin{figure}[htb]
\vspace*{-2.0cm}
\hspace*{-0.5cm}
\centerline{
\epsfxsize=7.0cm\epsfysize=9.0cm
\epsfbox{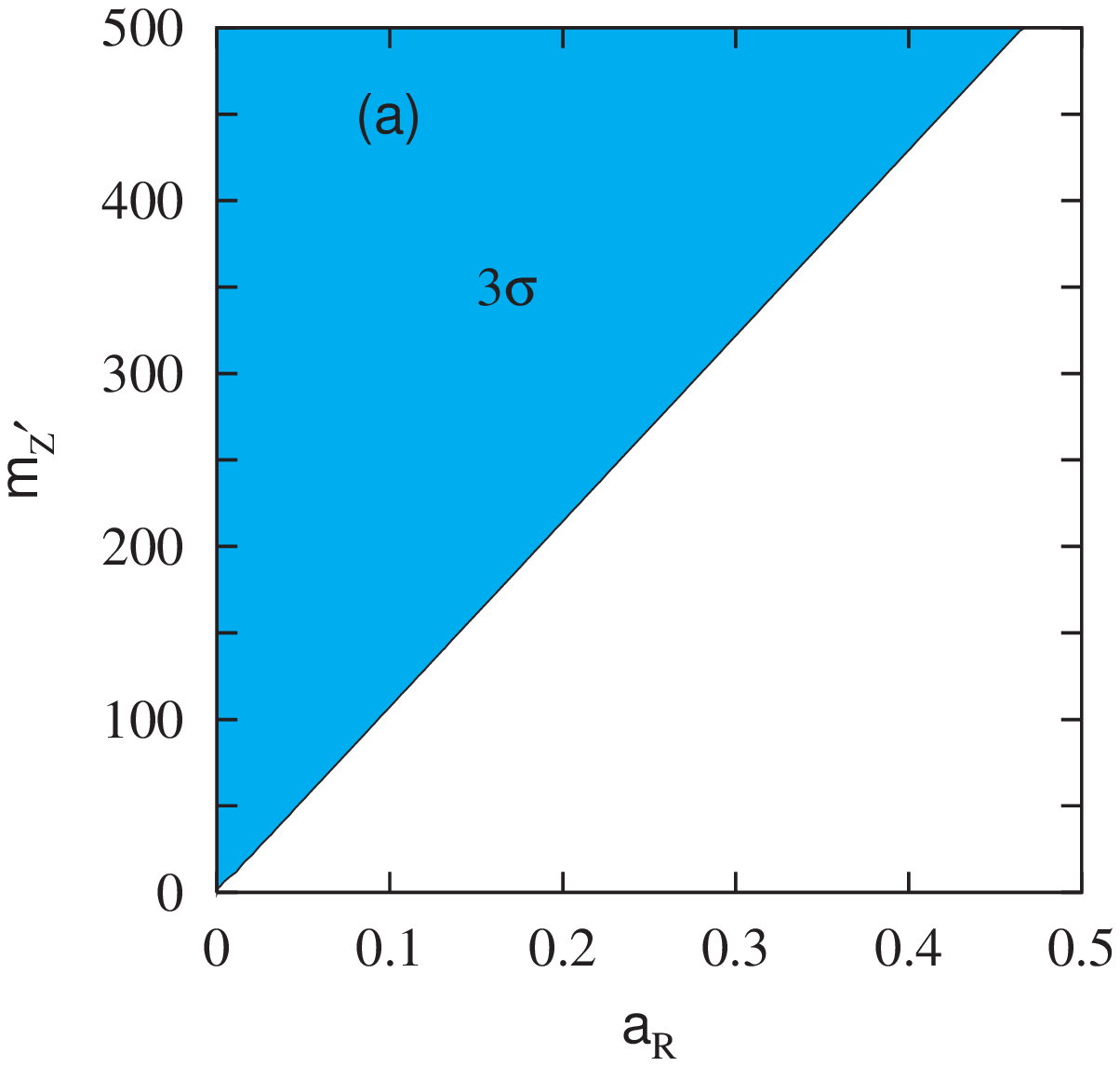} 
\vspace*{-0.0cm}
\hspace*{-0.cm}
\epsfxsize=7.0cm\epsfysize=9.0cm
\epsfbox{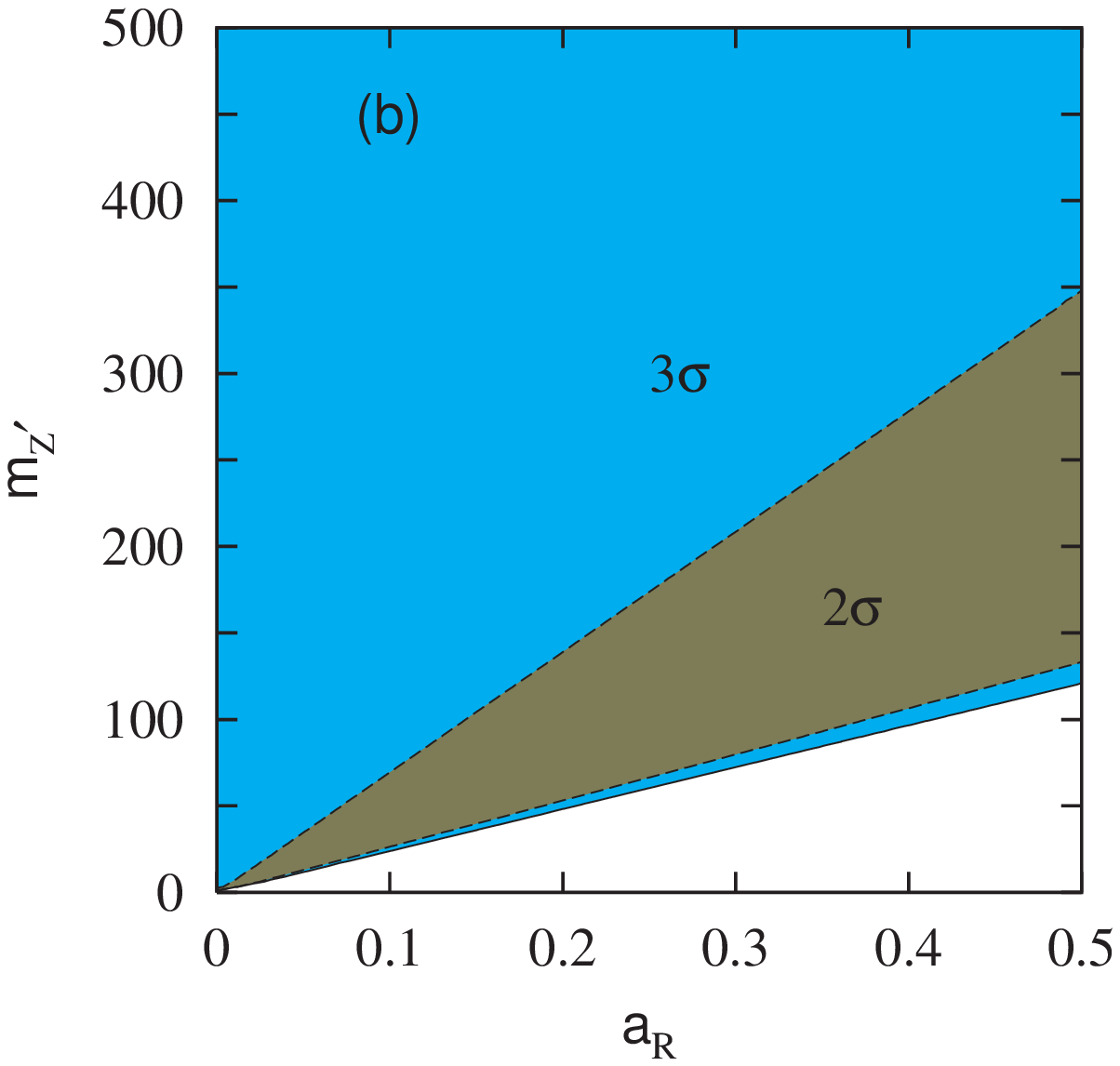}
\vspace*{-2.0cm}
}
\caption{\em The region of parameter space allowed by the $a_\mu$ data at
  different confidence levels for models with an extra $Z'$.  The
  light-shaded region is allowed at the $3\sigma$ level, whereas the
  darker one, to $2\sigma$ level.  Fig.~{\em (a)} corresponds to
  $a_L=-a_R$, and while Fig.~{\em (b)}, to $a_L=a_R$. Case~{\em (a)} is
  inconsistent with the data at the $2\sigma$ level.}
      \label{extraz}
\end{figure}

For the sake of convenience, we assume the gauge coupling associated
with the $Z'$ to be the same as that for the $Z$ and absorb any
deviation thereof into $a_{L,R}$. These two couplings, along with
$M_{Z'}$, then describe the theory.  To reduce the number of
parameters, we examine again cases with $a_L=a_R$ and $a_L=-a_R$.
Fig.~\ref{extraz} shows the area in the $a_R$ -- $M_{Z'}$ plane
excluded by the $a_\mu$ data for each of these two cases.  
Once again, the term proportional to $\frac{m_\tau}{M_{Z'}}$ dominates
the new contribution. Note, however, the considerable difference from
the case of the exotic lepton. The ratio $\frac{m_\tau}{M_{Z'}}$,
being much smaller than the typical value of $\frac{m_F}{M_Z}$ that we
considered, results in much smaller contribution in this case. This
immediately translates to a relaxation (see Fig. 3) on the bounds on
$a_{L,R}$ by a factor $\sim\big(\frac{m_F M_{Z'}}{m_\tau M_Z}\big)^{1/2}$.
Corrections to this naive estimate arise from the difference in values
of the function $f_2(r)$ in the two cases.

More importantly, the theory with an extra $Z'$ is a {\em decoupling
  one}. This has the immediate consequence of the bounds relaxing as
$M_{Z'}$ grows. In other words, the $(g_\mu-2)$ measurement is
progressively insensitive to a very heavy $Z'$, quite unlike the case
of an exotic lepton.

In conclusion, we have explored the phenomenological consequences of the
models containing extra gauge bosons or exotic fermions in the light of the
recent data on the anomalous magnetic moment of the muon from the E821
experiment at BNL.  We find that the constraints on the model parameters can
be significant if the flavour-changing couplings for these scenarios are
dominantly of the axial vector type. While the parameter space of vectorlike
fermion mixing with the muon gets rather strongly restricted by the recent
data, somewhat weaker but non-negligible constraints also arise for a $Z'$ if
the latter has flavour-changing interactions in the leptonic sector.

%
%

\end{document}